\PassOptionsToPackage{driverfallback=auto}{hyperref} 

\documentclass[fleqn]{2026SCGE}
\usepackage{bm} 
\usepackage{enumitem}  
\usepackage{xcolor}
\usepackage{stfloats}  
\usepackage{subcaption}  
\usepackage{hyperref} 
   
\renewcommand{\url}[1]{\href{https://doi.org/#1}{doi:\,#1}}
\begin{document}
\ensubject{subject}

\ArticleType{Article}
\Year{}
\Month{}
\Vol{}
\No{}
\DOI{}
\ArtNo{}
\ReceiveDate{}
\AcceptDate{}
\OnlineDate{} 

\title{Near-Horizon Tidal Disruption Events}{Near-Horizon Tidal Disruption Events}

\author[1,2]{Yang Yang}{}%
\author[1,2]{Xian Chen}{{xian.chen@pku.edu.cn}}

\AuthorMark{Yang Y}

\AuthorCitation{Yang Y, Chen X}

\address[1]{Department of Astronomy, School of Physics, Peking University, 100871 Beijing, China}
\address[2]{Kavli Institute for Astronomy and Astrophysics, Peking University, 100871 Beijing, China}


\abstract{
Tidal disruption events (TDEs) offer a unique dynamical probe of the spacetime geometry around supermassive black holes (SMBHs). While conventional TDEs occur around SMBHs of $M_\bullet\sim10^6$--$10^7M_\odot$, where stars are disrupted far from the event horizon, here we identify a special class of TDEs around rapidly spinning SMBHs with masses $M_\bullet\gtrsim10^8M_\odot$, where the tidal-disruption radius approaches the gravitational radius. We term these events ``near-horizon TDEs'' and, by calculating geodesics in Kerr spacetime, investigate how the proximity of the horizon modifies the debris evolution and subsequent fallback. We find that for stars encountering the SMBH on parabolic orbits, up to $\sim(90\%-95\%)$ of the stellar debris either plunges directly into the SMBH or escapes the system. Bound orbits, by contrast, retain a substantially larger fraction of the debris. Using the resulting debris distribution, we calculate the fallback rates and find that bound orbits produce intrinsically higher peak fallback rates and shorter decay timescales, by factors of $\sim10^3$ relative to parabolic orbits. However, when the stellar orbital angular momentum is particularly low, the peak fallback rate can be substantially suppressed by debris lost to plunge orbits. This combination of rapid fallback and a mass deficit naturally explains overluminous TDEs such as ASASSN-15lh, which standard TDE models have struggled to reproduce. Our work establishes near-horizon TDEs as a new probe of strong-field gravity and a promising tool for identifying massive, rapidly spinning SMBHs.  
}


\maketitle


\begin{multicols}{2}

\section{Introduction} 
\label{sec:1}  

Matter around black holes (BHs) provides a unique diagnostic for strong-field
gravity because both its dynamics and the associated electromagnetic (EM)
radiation encode the properties of the underlying spacetime
geometry~\cite{Bardeen1972,Misner1973,Chandrasekhar1983,Wald1984}.
Tidal disruption events (TDEs), in which a star is torn apart by a supermassive black hole (SMBH), have become a prime example of this diagnostic~\cite{Hills1975,Carter1982,Rees1988,Evans1989,Bade1996,Komossa1999}.  
More than a hundred TDEs have been observed~\cite{Komossa2015,Gezari2021,Langis2026} 
and some have been used to test relativistic effects, including the relativistic precession   ~\cite{Dai2015,Hayasaki2016,Jiang2016,Liu2017,Lu2020,Bonnerot2020,Chen2021,Pasham2024,Huang2024b,Yao2025}, relativistic line-profile distortions~\cite{Zhang2015,Liu2017,Roth2018,Mummery2020,Wevers2022b},
and  
the  spin-induced warping of the accretion disk~\cite{Lei2013,Shen2014,Franchini2016,Zanazzi2019,Pasham2024,Chen2026}.  \Authorfootnote
Theoretically, the most stringent test would be provided by a TDE in which the star is disrupted in the immediate vicinity of the event horizon~\cite{Kesden2012,Stone2019,Mummery2024,Huang2024,Banerjee2026}.

For a near-horizon TDE to happen, the SMBH must be sufficiently massive, since
more massive BHs have larger curvature radii (weaker tidal forces) near the
horizon, allowing the tidal-disruption radius ($r_t$)---where the tidal force exceeds the
stellar self-gravity---to approach the horizon scale.  
However, as the tidal radius approaches the event horizon, the orbit becomes increasingly unstable.
For a non-spinning (Schwarzschild) SMBH, the tidal radius of a solar-type star coincides with the event horizon at twice the gravitational radius $2r_g$\,$=$\,$2GM_\bullet/c^2$ when the BH mass $M_\bullet$ reaches
$\sim10^8 M_\odot$~\cite{Hills1975,Rees1988}. 
This tidal radius is significantly smaller than the innermost stable orbit (ISO)~\cite{Bardeen1972,Glampedakis2002,Levin2008}, which lies at $6r_g$ if the orbit is circular or $4r_g$ if the orbit is parabolic.  
Consequently, stellar debris from such a TDE would plunge directly into the SMBH with little opportunity to interact or produce an EM flare.  
Beyond this limit, as the BH mass increases further, the tidal radius moves inside the event horizon, making observable TDEs impossible.

However, SMBH spin mitigates this limitation by 
(i) allowing stable prograde orbits to exist closer to the horizon, and 
(ii) modifying the tidal tensor, thereby increasing the maximum BH mass $M_{\bullet,\max}$ for which a star can be disrupted outside the event horizon. 
For near-extremal spins and solar-type stars, this limit
increases to $M_{\bullet,\max}$\,$\sim$\,$10^9 M_\odot$~\cite{Kesden2012,Mummery2024,Xin2026}. 
Candidate events with SMBH  masses near or above this limit~\cite{Dong2016,Leloudas2016,Kruhler2018,Mummery2020,Hammerstein2023,Yao2025,Yao2026} may therefore indicate near-horizon TDEs happening around highly spinning
SMBHs, where Newtonian tidal-disruption models are invalid and could lead to
misleading observational interpretations.

On the theoretical front, the relativistic disruption dynamics of TDEs around SMBHs with $M_\bullet$\,$\gtrsim$\,$10^8 M_\odot$ remain largely unexplored. 
Previous general-relativistic hydrodynamic simulations of main-sequence star (MS) TDEs have
primarily focused on SMBHs with $M_\bullet$\,$\sim$\,$10^6\,$--$\,10^7 M_\odot$,  
for which the tidal radius lies well outside the strong-field region ($r_t$\,$\gg $\,$r_g$)~\cite{Laguna1993,Kobayashi2004,Evans2015,Hayasaki2016,Bonnerot2016,Skadowski2016,Tejeda2017,Darbha2019,Gafton2019,Liptai2019,Ryu2023,Chan2026,Calderon2026}.   
In these simulations, the initial disruption of the star occurs outside the strong-field region, and relativistic effects mainly influence the subsequent motion of the stellar debris rather than the initial disruption process.
In addition, although a few relativistic simulations have explored extremely deep encounters 
where the stellar orbital pericenter $r_p$ is significantly smaller than the tidal radius $r_t$~\cite{Kobayashi2004,Kesden2012b,Tejeda2017,Gafton2019,Ryu2023,Calderon2026},
their pericenters still remain at $\gtrsim 4r_g$.
Thus, the near-horizon regime with $r_t$\,$\sim$\,$r_p$\,$\sim$\,$r_g$ around highly spinning SMBHs remains poorly understood.

\begin{figure*}[!]  
    \centering 
    \setlength{\abovecaptionskip}{5pt}
    \begin{minipage}{\columnwidth}
        \centering
        \makebox[\columnwidth][r]{  
            \includegraphics[width=1.05\linewidth]{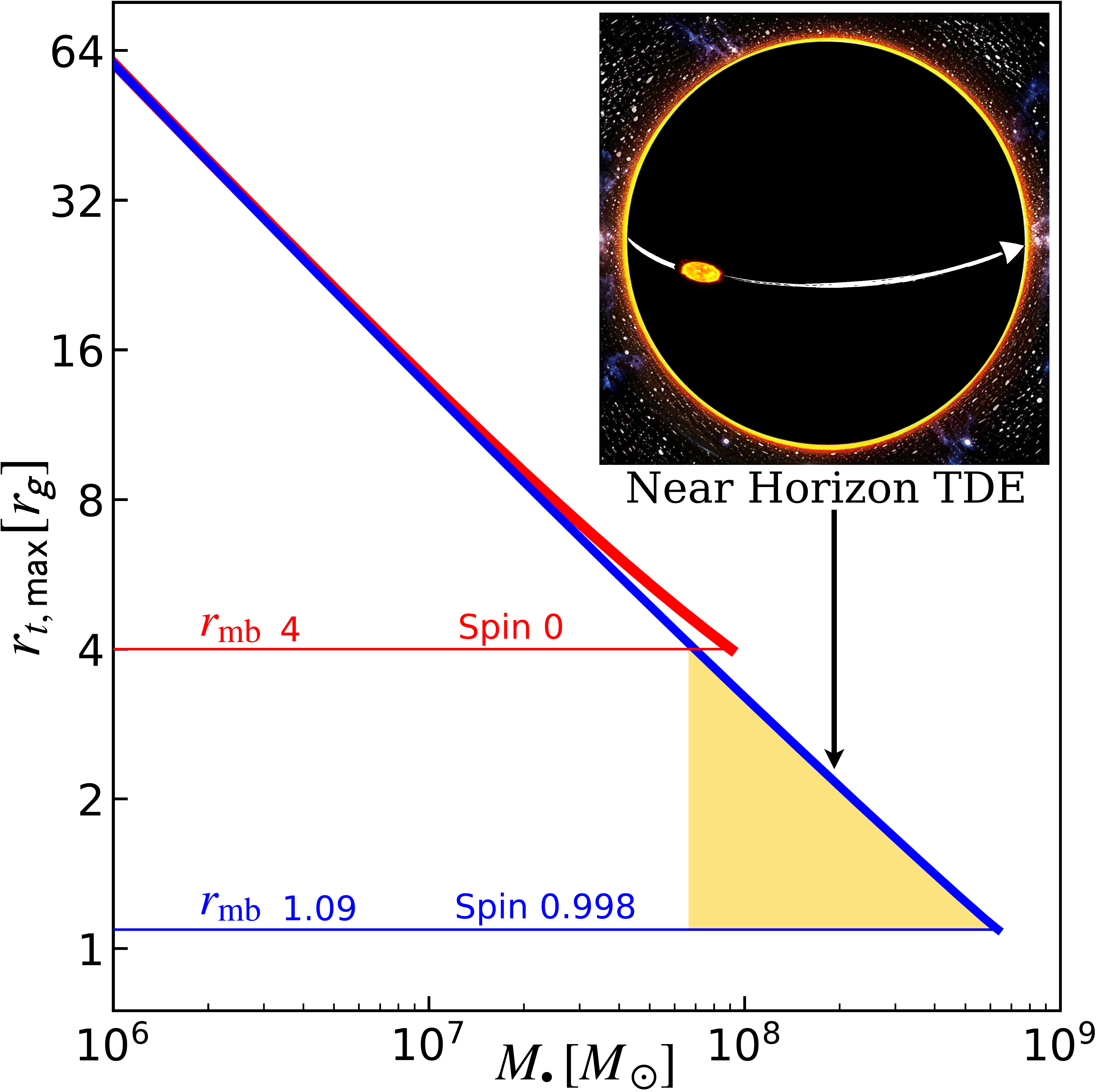} 
        }  
        \captionof{figure}{
        Maximum tidal radius $r_{t,{\rm max}}$ for a solar-type star ($M_{\star}$\,$=$\,$1M_\odot$) on parabolic equatorial geodesics around SMBHs with different masses. Red and blue curves correspond to non-spinning ($a_\bullet$\,$=$\,$0$) and near-extremal ($a_\bullet$\,$=$\,$0.998$) SMBHs, respectively. The horizontal lines show the minimum pericenter distances ($r_{\rm mb}$) set by marginally-bound orbits. The yellow shaded region marks our near-horizon TDE parameter space, with the inset showing illustrative artwork of a star much smaller than the high-mass SMBH near its horizon.
        }
        \label{fig:1}
    \end{minipage}  
    \hfill   
    \begin{minipage}{\columnwidth}
        \includegraphics[width= 0.955 \linewidth]{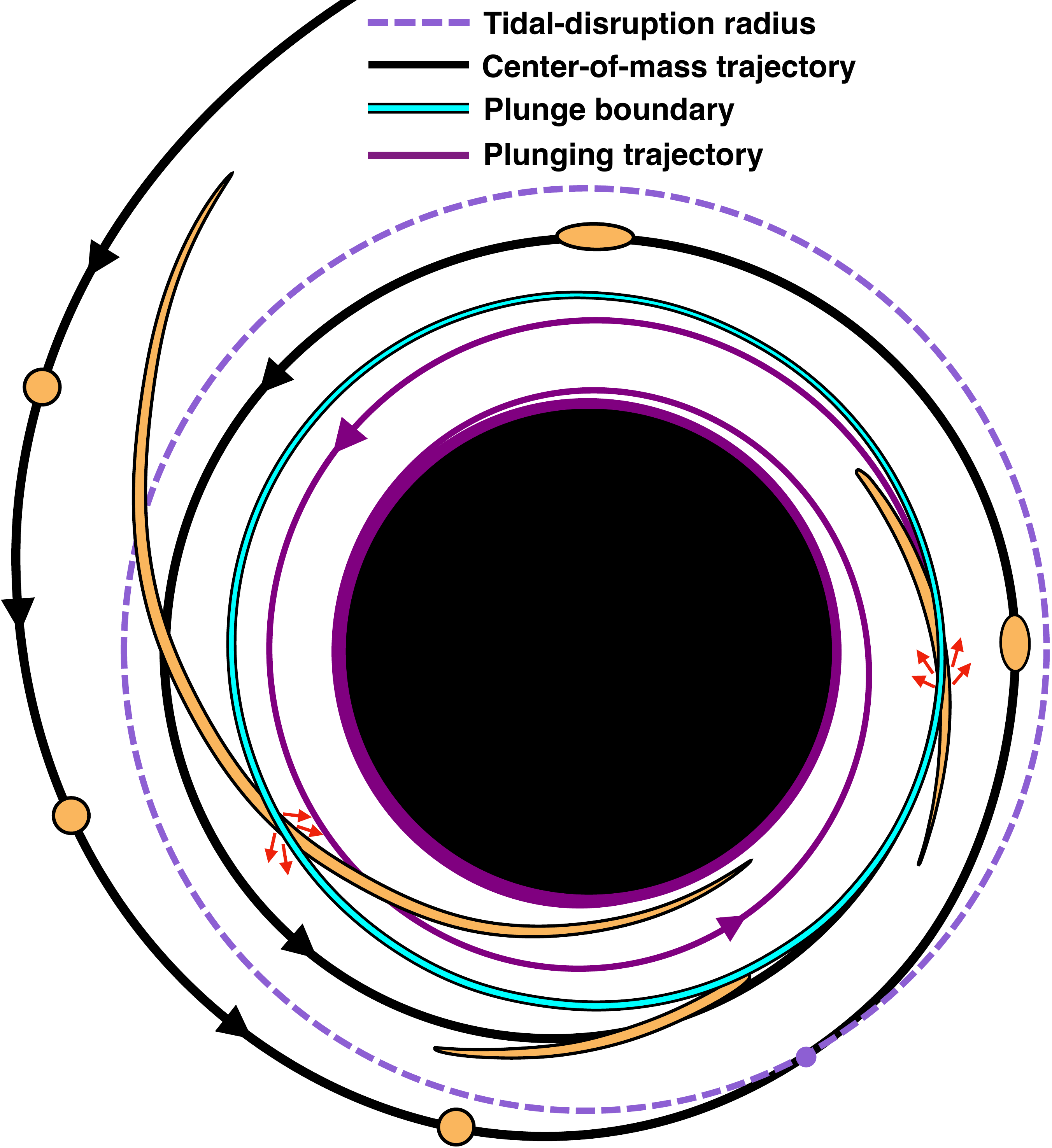}  
        \captionof{figure}{
        Schematic illustration of the early debris evolution of a near-horizon TDE around a highly spinning SMBH. 
        The light purple dashed circle marks the tidal radius  $r_t$. 
        The black solid curve shows the stellar CM trajectory, while the dark purple solid curve represents a plunging debris trajectory. 
        The cyan solid circle indicates the boundary separating plunging and non-plunging debris trajectories.  
        The red arrows indicate the separation of stellar debris at the above plunge boundary. 
        Notice that the figure is not drawn to scale. }
        \label{fig:2}
    \end{minipage}
    \vspace*{-10pt} 
\end{figure*}

Near the event horizon, stellar disruptions can exhibit behaviors qualitatively
different from those conventional TDEs.  
First, the critical eccentricity separating fully bound from partially unbound debris approaches unity as the SMBH mass approaches $M_{\bullet,\max}$~\cite{Hayasaki2018,Zhong2023,Park2020,Cufari2022b,Liu2023}.  
This implies that the conventional assumption of a parabolic stellar orbit may become inadequate in this regime.  
Consequently, the fraction of bound debris and hence the fallback timescale can be sensitive to small variations in the orbital parameters of the disrupted star. 
Second, the debris distribution in the
energy-angular momentum space can overlap with the parameter space occupied by
plunge orbits, causing a fraction of the debris to fall directly into the SMBH
immediately after tidal
disruption~\cite{Kobayashi2004,Kesden2012b,Tejeda2017,Ryu2023}. This removal of
debris from the fallback stream may produce a fallback rate that deviates from
the conventional symmetric bound-unbound
distribution~\cite{Rees1988,Lodato2009}. 
However, a systematic
comparison of the fallback rates in near-horizon and conventional TDEs has not yet been carried out.

This work aims to identify the SMBH systems relevant for near-horizon TDEs,
study the resulting debris dynamics, and characterize the modified fallback rates. 
The paper is organized as follows. 
In sect.~\ref{sec:2}, we define the near-horizon TDE regime. 
In sect.~\ref{sec:3}, we develop the theoretical framework for plunge-induced mass deficits and calculate the corresponding fallback evolution. 
In sect.~\ref{sec:4}, we discuss observational implications, including the application of our model to ASASSN-15lh and other near-horizon
TDE candidates. 
Finally, we summarize our conclusions in sect.~\ref{sec:5}.

\section{The Importance of Spinning SMBHs} \label{sec:2} 

If a star is tidally disrupted in the vicinity of the event horizon of an SMBH,
we refer to this event as a near-horizon TDE. The case is more interesting if
the SMBH is highly spinning because the event horizon is smaller than the
Schwarzschild case, the relativistic effects are stronger, and the
tidal radius could become comparable to the gravitational radius,
\textit{i.e.}, $r_t\sim r_g$.

To find out where $r_t$ is, 
it is important to realize that close to the event horizon of an SMBH, 
$r_t$ depends not only on the masses of the star and the
SMBH, but also on the orbital energy and angular momentum of the
star~\cite{Kesden2012,Rossi2021,Mummery2024,Xin2026}.    
The tidal radius is determined by evaluating the relativistic tidal tensor along the stellar center-of-mass (CM) geodesic and locating the radius where its largest eigenvalue becomes comparable to the star's self-gravity~\cite{Rossi2021}.
In particular, given the specific energy $E$ of the star, there is a critical
specific angular momentum $L_{\rm TDE}(E)$   
at which 
TDE starts to happen. When $L=L_{\rm TDE}$, the star is disrupted
at the orbital pericenter, \textit{i.e.}, the tidal radius $r_t$ coincides with the pericenter
distance $r_p$. Below $L_{\rm TDE}$, both $r_t$ and $r_p$ get smaller as  $L$ decreases, but 
$r_t$ always stays larger than $r_p$.

In this sense, there is a maximum tidal radius $r_{t,\rm max}$ for each $E$, which is equal to the pericenter distance at $L=L_{\rm TDE}$. 
For example, for a solar-type  star ($M_{\star}$\,$=$\,$1M_\odot$) with $E/c^2$\,$=$\,$1$, 
\textit{i.e.},  the orbit is parabolic,
the location of $r_{t,\rm max}$ and its dependence on SMBH mass are shown in Figure~\ref{fig:1}. 
Here we have considered both a non-spinning
Schwarzschild SMBH (thick red curve) and a  highly spinning SMBH  with a dimensionless spin parameter 
$a_\bullet$\,$=$\,$0.998$ (thick blue curve).  
In both cases, $r_{t,\rm max}$ is getting closer to $r_g$ as the BH mass increases,
indicating stronger relativistic effects.  
The stronger relativistic effects also render the two $r_{t,\rm max}$ curves more different, highlighting the necessity of this work. 
Figure~\ref{fig:1} also shows that the $r_{t,\rm max}$ curves truncate at 
$r_{\rm mb}$, since the minimum pericenter distance is set by the marginally-bound orbit 
(MBO)~\cite{Bardeen1972}.  
In this work, we do not consider cases where the tidal radius $r_t$ lies inside  $r_{\rm mb}$, 
because the stellar orbit would be in the plunging regime.

The TDEs of our interest reside below the $r_{t,\max}$ curve and
above the $r_{\rm mb}$ line. As Figure~\ref{fig:1} suggests, for parabolic
stellar orbits, this region stays above $4r_g$ when the SMBH is non-spinning.
This is twice the horizon size of a Schwarzschild BH. When $a_\bullet=0.998$,
however, the region can penetrate $4r_g$ and almost reach the horizon at
$1r_g$ if the SMBH is between $M_\bullet\sim7\times10^7M_\odot$ and
$7\times10^8M_\odot$ (yellow shaded region). 
This is the reason why we will focus on highly spinning SMBHs in the following analysis.

In the near-horizon regime, the stellar CM orbit lies close to the
plunge boundary, giving rise to two additional effects that fundamentally
alter the post-disruption evolution compared with conventional TDEs (see Figure~\ref{fig:2}). First,
such trajectories can undergo a relativistic whirl phase~\cite{Glampedakis2002,Levin2008,Mummery2023}, during which the star
completes multiple azimuthal
cycles slightly above the plunge boundary before either escaping or
plunging. 
The prolonged residence in the strong-field regime produces extreme tidal deformation and spiral-like
debris structures~\cite{Laguna1993,Kobayashi2004,Tejeda2017,Gafton2019,Ryu2023}, while the extreme apsidal precession (can exceed $2\pi$) with different precession angles
for different debris elements renders conventional single-stream
self-intersection circularization
models~\cite{Dai2013,Dai2015,Guillochon2015,Hayasaki2016,Jiang2016,Lu2020,Chen2021,Batra2023}
inapplicable. 
Instead, returning and infalling streams continuously interact
during the whirl phase, enabling rapid circularization within only a few azimuthal cycles~\cite{Leloudas2016}. 
Quantitatively capturing this circularization process, however, requires detailed hydrodynamic simulations of the stream–stream interactions during the whirl phase, which we leave for future work.

Second, due to the finite size of the star, 
the debris elements acquire a spread in
orbital energy and angular momentum~\cite{Kesden2012b,Stone2019}. 
Those with
angular momentum above the plunge boundary can avoid direct plunge and
may return (fallback) to the SMBH  if their specific binding energy exceeds the rest-mass energy in magnitude (\textit{i.e.}, they are gravitationally bound), 
whereas those crossing the plunge boundary directly fall into the SMBH during the first passage. 
The latter plunge produces an asymmetric
truncation of the debris distribution that tears the stream apart and results
in a mass deficit in the fallback material~\cite{Kobayashi2004,Kesden2012b,Tejeda2017,Ryu2023}.  
How this mass-deficit effect would affect TDE light curves and offer observational hints will be studied in detail in the following two sections.

\section{Near-Horizon TDEs} \label{sec:3} 
 
\subsection{Relevant Stellar Orbits} \label{sec:3.1}

Now we have understood what orbits correspond to near-horizon TDEs.
To understand the subsequent evolution of the stellar debris, we
revisit the geodesics in a Kerr metric~\cite{Kerr1963}.  
We adopt the
Boyer--Lindquist coordinates (BLCs)
$x^\mu$\,$=$\,$\left ( t,r,\theta,\phi \right )$~\cite{Bardeen1972,Chandrasekhar1983}, and use three
conserved quantities to characterize a geodesic: the specific energy $E$, the
axial component of the specific angular momentum $L_z$, and the Carter constant
$Q$~\cite{Carter1968,Wald1984}.  
We focus on equatorial orbits partly because
they are simpler, \textit{e.g.}, choosing $Q$\,$=$\,$0$ reduces the parameter space $(E,L_z,Q)$ to a
two-dimensional $(E,L$\,$=$\,$L_z)$ space where the plunge boundary can be more clearly
characterized~\cite{Bardeen1972,Chandrasekhar1983,Glampedakis2002}. More
importantly, close to the equator, the spread in the Carter constant $Q$
remains a second-order correction, $\delta
Q\propto(\delta\theta)^2$~\cite{Carter1968,Kesden2012b}, so that to leading
order the distribution of stellar debris can be described without explicitly
including $Q$.

\vspace{-10pt}
\begin{figure}[H]
\centering
\setlength{\abovecaptionskip}{3pt}
\makebox[\columnwidth][r]{  
    \includegraphics[width=1.062\columnwidth]{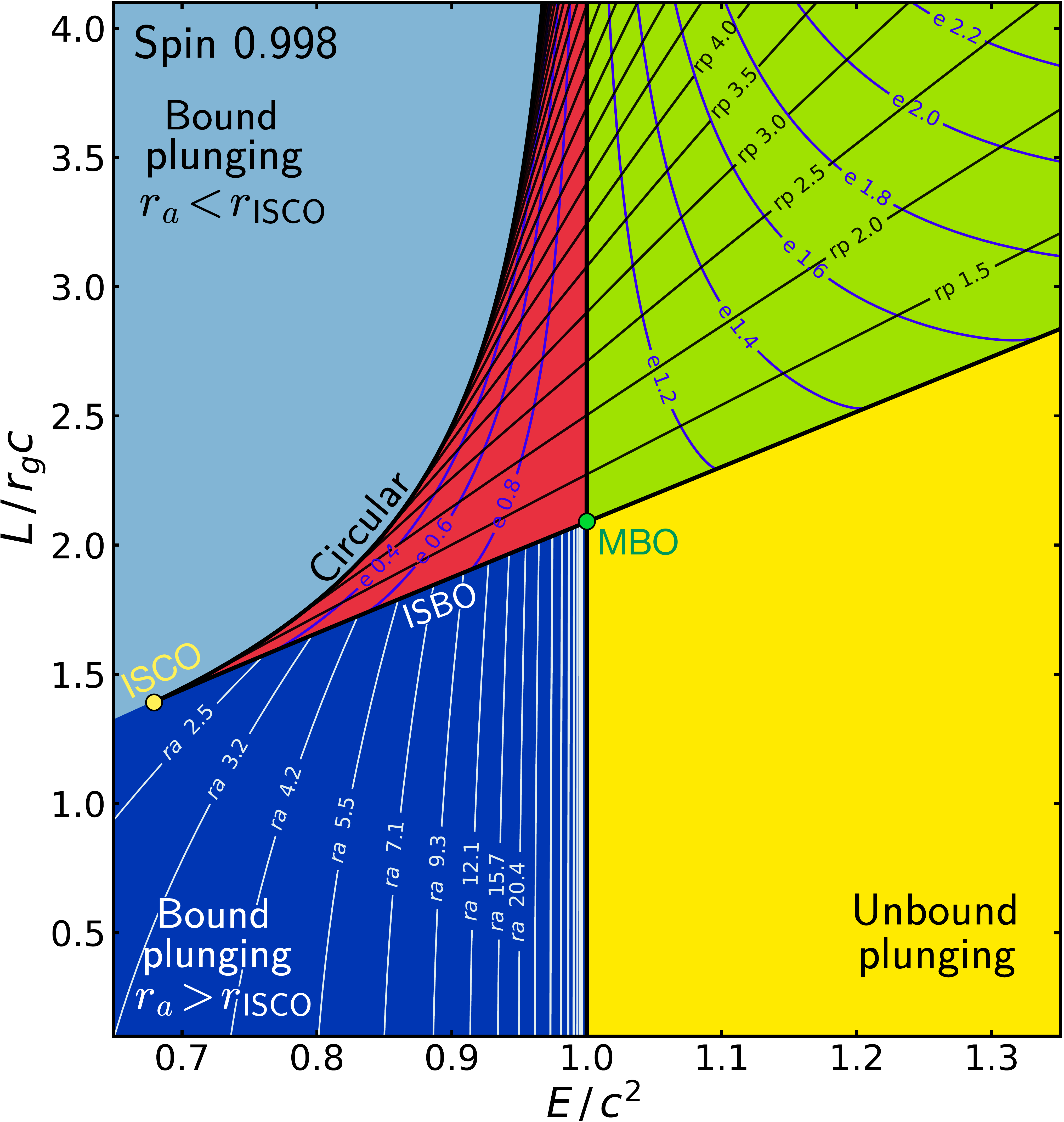}%
}
\caption{
    Equatorial orbital types around a near-extremal BH ($a_{\bullet}=0.998$).
    Contour lines show the apocenter distance $r_a$ (white), pericenter distance $r_p$ (black), and eccentricity $e$ (blue).
    The color regions denote different orbital classes:
    green: unbound escaping orbits;
    yellow: unbound plunging orbits;
    red: bound nonplunging orbits;
    blue: bound plunging orbits with $r_a$\,$>$\,$r_{\rm ISCO}$;
    slate-blue: bound plunging orbits with $r_a$\,$<$\,$r_{\rm ISCO}$.
    The yellow region does not show contours because the geodesics here
    originate from infinity and plunge directly into the BH without reaching a radial turning point.
    The slate-blue region contains geodesics which are confined inside the ISCO and will fall into the event horizon, therefore will not be further considered in this work.
}
\label{fig:3}
\end{figure}

For an equatorial geodesic, the orbital type is determined
by the structure of the radial potential, more specifically,
the number and properties of the
roots~\cite{Bardeen1972,Chandrasekhar1983,Glampedakis2002,Levin2008,Levin2009,Mummery2024}.
Figure~\ref{fig:3} shows the orbital classification in the $E-L$ plane around an SMBH
with $a_\bullet=0.998$.  The vertical line at $E/c^2=1$ divides the parameter space into bound and unbound sections, and bound orbits lie to the left of this line. 
The thick line passing through the point labeled ``MBO''  (marginally-bound orbit) marks the
plunge boundary, and below this line test particles will plunge into the BH. 
Circular equatorial orbits form a 
characteristic curve labeled ``Circular'' in the $E-L$ plane. 
This curve intersects the plunge boundary at the point labeled 
the ``innermost stable circular orbit'' (ISCO).
The line segment between ISCO and MBO is occupied by bound but marginally stable orbits, which
we call ``the innermost stable bound orbit'' (ISBO).

Near-horizon TDEs happen in a limited region in the $E-L$ plane.  First, the CM
of the star must approach the horizon as much as possible, but it should remain
close  to the plunge boundary so that some stellar debris has a chance to mutually
interact and produce EM radiation. 
Second, we have mentioned in sect.~\ref{sec:1}
that the energy spread of the stellar debris is much smaller than  the CM orbital energy $E$. 
If $E$ initially is greater than $c^2$, all the stellar debris may have $E/c^2>1$ immediately after tidal disruption and will escape the system without interaction. 
Therefore, to produce EM signals, near-horizon TDEs should lie
close to the $E/c^2=1$ line, or to the left of it.

These two requirements, \textit{i.e.}, proximity to the plunge boundary and $E/c^2\lesssim 1$,
naturally select the ISBO branch 
as the relevant parameter space for near-horizon TDEs. In the following, we first examine the 
limiting case where the CM follows the MBO, the rightmost point of the ISBO branch.

\subsection{Distribution of Debris in the $E-L$ plane} \label{sec:3.2}

Upon disruption at the tidal radius $r_t$ (normally $>r_p$), different parts of
a star have different $E$ and $L$ due to their displacements from the CM of
the star.  As long as the displacements are small relative to the curvature
radius of the background, which is true in our problem because $R_\star$\,$\ll
$\,$r_g$, the deviations $\Delta E$ and $\Delta L$ from the CM energy $E_c$ and CM
angular momentum $L_c$ can be calculated in a Fermi normal coordinate (FNC)
system~\cite{Manasse1963,Marck1983,Ishii2005,Klein2008,Kesden2012b,Xin2026,Yang2026}.  The deviation of Carter constant is neglected
because our TDEs happen in the equatorial plane (see sect.~\ref{sec:3.1}).  

Following ref.~\cite{Kesden2012b}, we write
\begin{equation}\label{eq: DeltaEL formula}
\Delta E  =  - \Gamma_{\alpha t}^{\gamma} g_{\gamma\beta} U_{\rm c}{}^{\beta} \lambda^{ \alpha }{ }_{I}X^I,
\qquad
\Delta L  =  \Gamma_{\alpha \phi}^{\gamma} g_{\gamma\beta} U_{\rm c}{}^{\beta} \lambda^{ \alpha }{ }_{I}X^I .
\end{equation}
Here, the indices $\alpha,\beta,\gamma$ denote the BLCs and $I$ denotes the
spatial directions in the FNCs.  The four-velocity $U_{\rm c}{}^{\beta}$ refers
to the CM of the star, $\lambda^{ \alpha }{ }_{I}$ denotes the spatial tetrads
parallel-transported along the CM geodesic~\cite{Marck1983,Kesden2012b}, and
$X^I$ is the displacement of a fluid element from the stellar CM in the FNCs.
The Kerr metric $g_{\gamma\beta}$ and Christoffel symbols 
$\Gamma^\gamma_{\alpha\beta}$ are evaluated at $r=r_t$ and  treated as constants across
the FNC (\textit{i.e.}, independent of
$X^I$)~\cite{Misner1973,Wald1984,Kesden2012b}.  Equation~\eqref{eq: DeltaEL
formula} indicates that $\Delta E$ and $\Delta L$ differ only in the Killing direction with which they are associated: $\Gamma^\gamma_{\alpha
t}$ corresponds to the time-translation symmetry, whereas $\Gamma^\gamma_{\alpha\phi}$ corresponds to the axial symmetry of the Kerr spacetime~\cite{Carter1968}.

Equation~\eqref{eq: DeltaEL formula} enables a linear mapping of the 
projected surface  density of the star
into the $E-L$ plane, which we now elaborate.
We first rotate the FNC displacement vector $\boldsymbol{X}$\,$=$\,$(X^1,X^2,X^3)$ and  rewrite the equation as
\begin{equation}\label{eq: DeltaEL formula A B}
\Delta E=\boldsymbol{A}\cdot\boldsymbol{\xi},\qquad
\Delta L=\boldsymbol{B}\cdot\boldsymbol{\xi},
\end{equation}
where $\boldsymbol{A}$ and $\boldsymbol{B}$ are determined by the coefficients of
$\boldsymbol{X}$ in eq.~\eqref{eq: DeltaEL formula}, and 
$\boldsymbol{\xi}=\xi_1 \,\boldsymbol{n}_1+\xi_2\,\boldsymbol{n}_2+\xi_3\,\boldsymbol{n}_3$
is the displacement vector $\boldsymbol{X}$ expressed in the rotated orthonormal basis $\{\boldsymbol{n}_1,\boldsymbol{n}_2,\boldsymbol{n}_3\}$ and normalized by $r_g$, \textit{i.e.}, $(\xi_1)^2$\,$+$\,$(\xi_2)^2$\,$+$\,$(\xi_3)^2$\,$\leq$\,$(R_\star/r_g)^2$. 
Here, the basis $\{\boldsymbol{n}_1,\boldsymbol{n}_2,\boldsymbol{n}_3\}$ is chosen such
that $(\boldsymbol{n}_1\parallel \boldsymbol{A})$ and $\boldsymbol{n}_2$ lies in the plane
spanned by $\boldsymbol{A}$ and $\boldsymbol{B}$.
Thus, eq.~\eqref{eq: DeltaEL formula A B} reduces to
\begin{equation}\label{eq: DeltaEL formula xi}
\Delta E=A\,\xi_1,\qquad
\Delta L=B_1\,\xi_1+B_2\,\xi_2 ,
\end{equation}
where $A=|\boldsymbol{A}|$, $B_1=\boldsymbol{B}\cdot\boldsymbol{n}_1$, and $B_2=\boldsymbol{B}\cdot\boldsymbol{n}_2$.

Since $\Delta E$ and $\Delta L$ depend only on $\xi_1$ and $\xi_2$, 
fluid elements sharing the same  $(\xi_1,\xi_2)$---regardless of their 
$\xi_3$---map to identical points in the $E-L$ plane. 
We can thus integrate over $\xi_3$ and express the mass distribution
in the $E-L$ plane via the projected surface density of the star.  
In particular, a cylindrical surface (constant $\xi_1^2+\xi_2^2$)  maps to an ellipse in the $\Delta E-\Delta L$ plane, given by 
\begin{equation}\label{eq: elliptical formula} 
     f(\Delta E,\Delta L) \equiv \left(\frac{\Delta E}{A}\right)^2+
    \left(
    \frac{\Delta L-B_1\, \Delta E/A}{B_2}
    \right)^2
    = \text{constant},
\end{equation}
with the constant determined by the cylindrical radius. 
The debris mass distribution on this ellipse is
\begin{equation}\label{eq: mass distribution formula }
\begin{aligned}
    &\ \frac{\mathrm{d}^{2} M}{\mathrm{d}\Delta E\, \mathrm{d}\Delta L}\left ( \Delta E ,\, \Delta L,\,E_c,\,L_c,\,r_t \right )     \\
    = &\int \! \! \mathrm{d}^3  \xi ~ \rho \left(\left | \boldsymbol{\xi}  \right | \right) 
    ~\delta \! \left(\Delta E - A \, \xi_1 \right) 
    ~\delta \! \left(\Delta L - B_1\,\xi_1 - B_2\,\xi_2\right)  \\
    =  & \ \frac{1}{A\left | B_2 \right | } \int \! \mathrm{d}\xi_3 \, \rho \left( 
    \sqrt{  (\xi_3)^2 + f(\Delta E,\Delta L) } 
    \right)  \\ 
    =  & \ \frac{1 }{A\left | B_2 \right | } \, \Sigma \left( 
    \sqrt{  f(\Delta E,\Delta L) } 
    \right),     
\end{aligned}
\end{equation}
where $\rho\left(\left | \boldsymbol{\xi}  \right | \right)$ is the radial density profile 
of the spherically symmetric star,  and  $\Sigma$ is the corresponding projected surface density.

According the above distribution, the half-width of the energy distribution is
\begin{equation}\label{eq: Delta E_max} 
    \Delta E_{\rm max}(E_c,\,L_c,\,r_t)/c^2=A\,R_\star/r_g.
\end{equation}
For MBOs, we use $\Delta E_{\rm MBO,max}$ to denote the half-width and use it as a unit for measuring the energy offset from the CM energy  $E_c$ when needed.

\begin{figure}[H] 
    \centering
    \setlength{\abovecaptionskip}{5pt}
    \vspace*{-5pt} 
    \makebox[\columnwidth][c]{  
        \includegraphics[width=1.08\columnwidth]{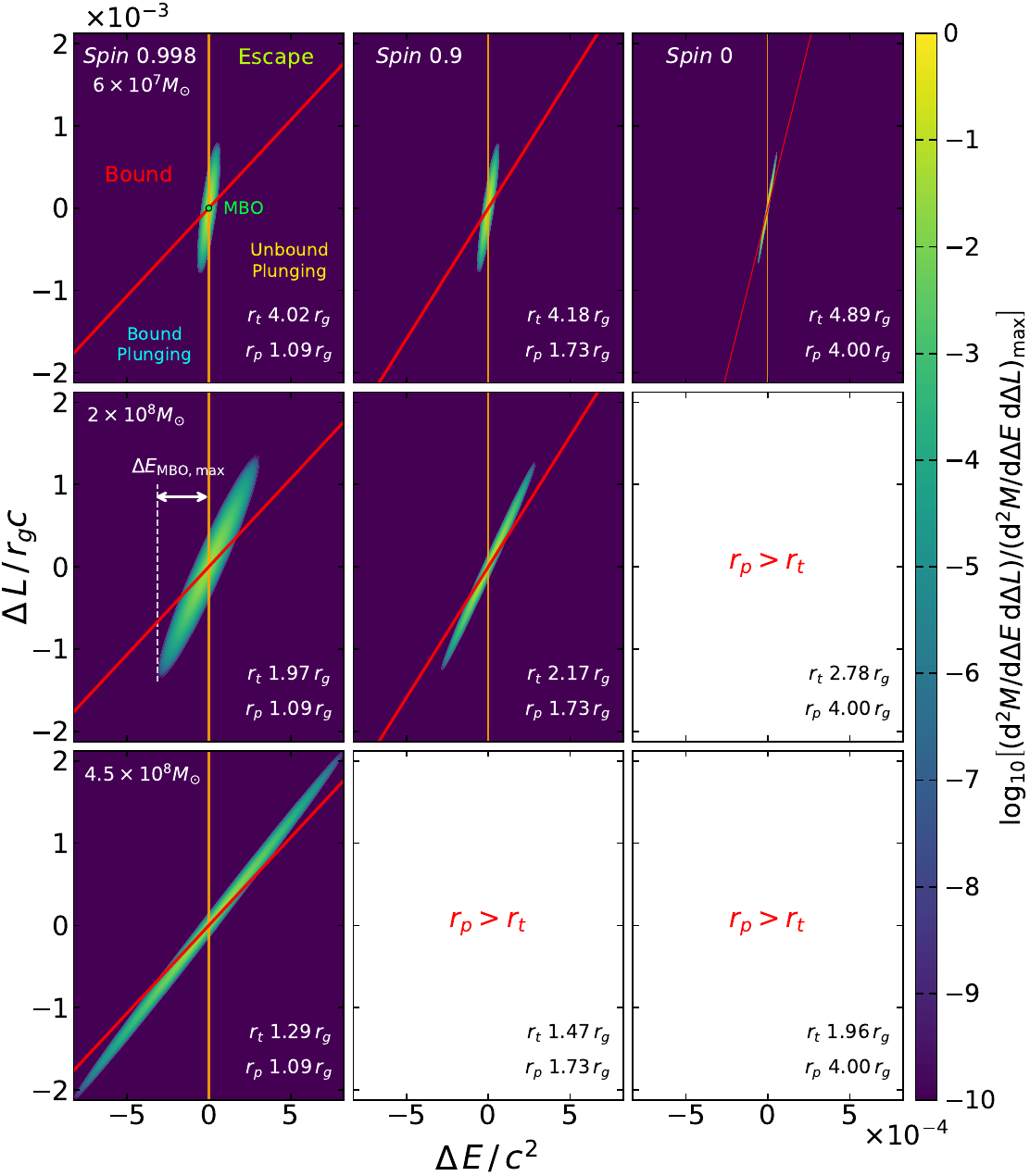}%
    } 
    \caption{
        Distribution of stellar debris in the $\Delta E-\Delta L$ plane. 
        The energy and angular momentum are evaluated at $r$\,$=$\,$r_t$, where we have assumed a solar-type star on an MBO.
    The color indicates the logarithmic mass density normalized to its maximum value, with the orange and  red lines indicating the escape and plunge boundaries, respectively.
    Panels in different rows correspond to different SMBH masses (increasing downwards), and different columns correspond to different spins (decreasing rightwards). 
    Blank panels correspond to the cases where $r_p$\,$>$\,$r_t$ --- complete disruption is difficult.
    }
    \label{fig:4} 
    \vspace*{-5pt}  
\end{figure}

\begin{figure*}
    \centering
    \captionsetup[subfloat]{position=top}
    \setlength{\abovecaptionskip}{5pt}
    \vspace*{-15pt}
    
    \makebox[\textwidth][c]{%
        \subfloat[]{
            \includegraphics[width=0.432\textwidth]{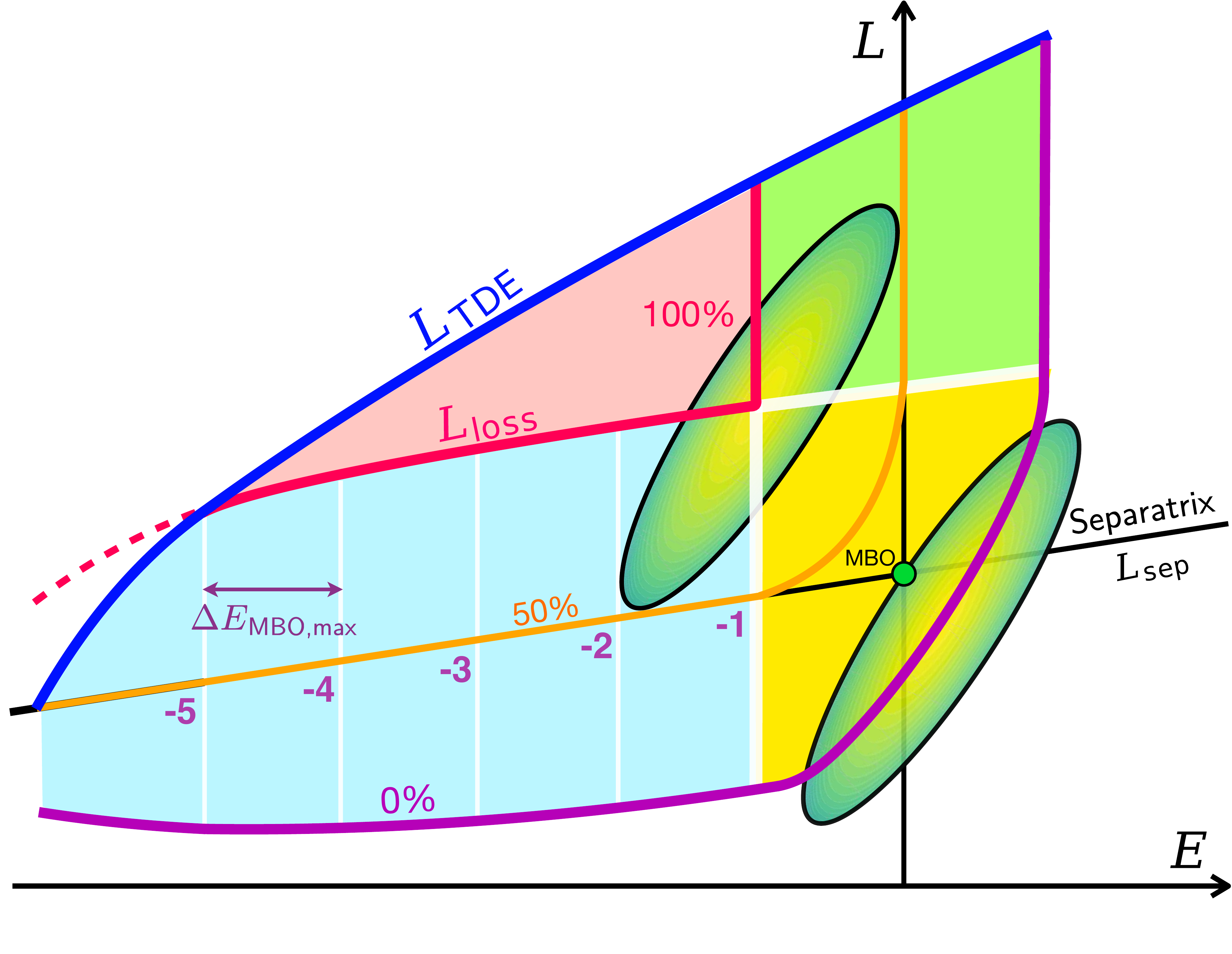}
            \label{fig:5a}
        }%
        \hfill
        \subfloat[]{
            \includegraphics[width=0.578\textwidth]{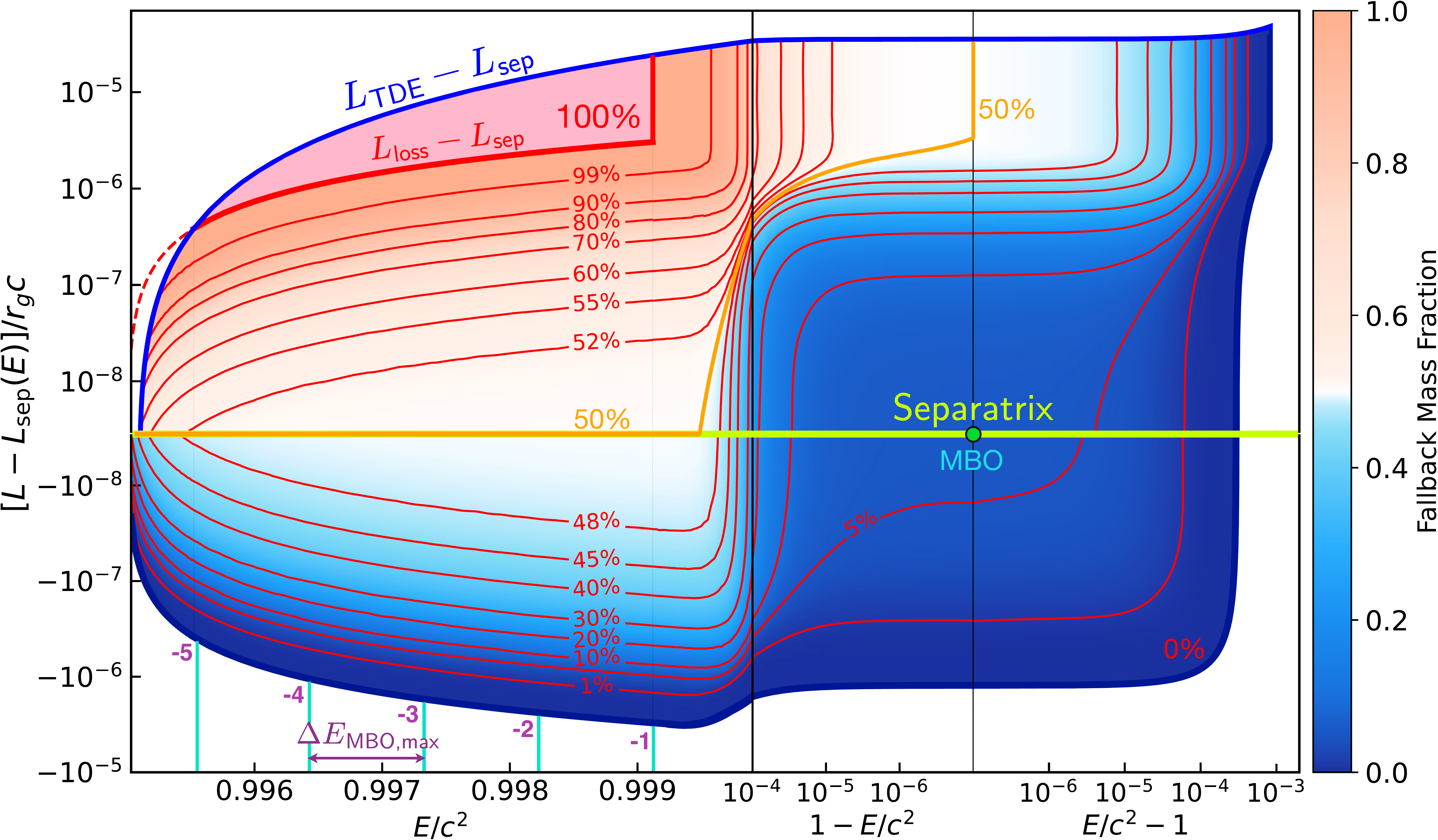}
            \label{fig:5b}
        }%
    } 
    \caption{
    Diagnostic plane for the amount of fallback mass.
    Panel (a) shows the distribution of stellar debris and the boundaries separating different outcomes in the $E-L$ plane.
    The oblique ellipses represent the debris spread, whose locations are determined by the CM orbital parameters $(E_c,L_c)$.  
    The blue curve marks the boundary for complete tidal disrupstion ($L_{\rm TDE}$),  the red curves mark the boundary above which there is no mass loss, and the purple curve separates  the region where the stellar mass is completely lost due to either plunging or escaping. 
    Accordingly, the red shaded region corresponds to complete fallback, the cyan region has mass loss due to plunge, the green one also loses mass due to escaping unbound debris, and both mass-loss mechanisms take effect in the yellow shaded region.
    The vertical white lines marked by $-5$ to $-1$ indicate the multiples of the energy half-width $\Delta E_{\rm MBO,max}$, defined in sect.~\ref{sec:3.2}.  
    Panel (b) is essentially the same as Panel (a) but calculated for a solar-type star ($n$\,$=$\,$3$), disrupted by an SMBH with  $M_\bullet$\,$=$\,$6.33\times10^8 M_{\odot}$ and $a_\bullet$\,$=$\,$0.998$, and plotted using carefully chosen coordinates to highlight the details in interesting parameter space. 
    The color corresponds the fallback mass fraction of the stellar debris that remains bound to the SMBH and avoids direct plunge into the horizon.
    These debris can produce luminous flares after the star is disrupted.  }
    \label{fig:5} 
    \vspace*{-5.5pt} 
\end{figure*}

Figure~\ref{fig:4} shows the debris mass distribution in the $\Delta E-\Delta L$ plane after a solar-type star on an MBO is disrupted.
The stellar structure is modeled as a Lane--Emden polytropic sphere, corresponding to an adiabatic index $\gamma$\,$=$\,$4/3$, with a polytropic index of $n$\,$=$\,$3$. 
This model provides an approximate radial density profile for a solar-type star~\cite{Chandrasekhar1939,Guillochon2013}.  
Rows and columns, respectively, correspond to different SMBH masses and spins. 
We adopt the same
orbital classification defined in sect.~\ref{sec:3.1} (Figure~\ref{fig:3}) to
study possible outcomes of the debris.  
Therefore, in each panel, the debris
is divided into four regions separated by the orange and the red demarcation
lines.  In the upper-left region, the debris is bound to the SMBH and, after a
brief outgoing phase, will fallback to the pericenter.  On the other hand, the
debris in the upper-right region is unbound, and will escape to infinity after
tidal disruption.  The two lower regions correspond to those debris that will
plunge into the SMBH immediately after the first pericenter passage.

The presence of debris in the lower-left region marks a fundamental departure
from Newtonian TDE models. In the conventional Newtonian scenario, all bound
material eventually returns to produce a luminous flare~\cite{Rees1988,Lodato2009,Park2020}. In the near-horizon
regime, however, we see that a fraction of the bound debris lies below the
plunge boundary and thus falls directly into the BH immediately after the first
pericenter passage. The plunging debris significantly reduces the mass budget
available for fallback and the subsequent emission. The observational implications
of this  plunge-induced
deficit, which has no Newtonian counterpart, will be discussed in
sect.~\ref{sec:4}.

Figure~\ref{fig:4} also shows that in the systems where $r_t$ is closer to
$r_p$, the distribution is more elongated in the $\Delta E-\Delta L$ plane.  In
the limit $r_t$\,$\sim$\,$r_p$\,$\sim$\,$r_g$ (lower-left panel), the distribution
stretches into an one-dimensional structure,  whose direction becomes nearly
parallel to the plunge boundary.  
As a result, the amount of debris that plunges into the SMBH becomes extremely sensitive to the orbital parameters of
the stellar CM.  
To understand how commonly a fallback mass deficit arises in near-horizon TDEs, we have to relax the assumption that the stellar CM follows an MBO and examine a broader range of CM orbital energies and angular momenta.

\subsection{Fallback Mass Deficits for arbitrary orbits} \label{sec:3.3} 

To study the fallback mass across a broader range of stellar CM orbits, it is useful to identify three critical boundaries in the $E-L$ plane, which separate different fallback regimes. They are shown in Figure~\ref{fig:5a}. 
\begin{description}[
    noitemsep,  
    topsep=3pt,  
    leftmargin=4.4em, 
    labelsep=0.5em,  
    align=left,   
    font=\hspace{1.2em}\textbullet\hspace{0.3em} 
]
    \item[\normalfont Blue:] Boundary for complete tidal disruption, \textit{i.e.}, 
    the angular-momentum threshold $L_{\rm TDE}(E)$ mentioned in sect.~\ref{sec:2}.
    Below this boundary, the star is completely disrupted.
    
    \item[\normalfont Red:] Boundary for fallback without mass loss, \textit{i.e.}, 
    the angular-momentum threshold $L_{\rm loss}(E)$ at which the debris ellipse 
    first intersects either the plunge boundary $L_{\rm sep}$ or the escape 
    boundary $E/c^2=1$. 
    Below this boundary, a fraction of the debris is removed 
    from the fallback material due to either direct plunge or unbound escape.

    \item[\normalfont Purple:] Boundary for complete mass loss, \textit{i.e.}, 
    the angular-momentum threshold at which the entire debris ellipse lies in 
    the plunge or escape region. 
    Below this boundary, all debris elements 
    either directly plunge into or escape from the SMBH, leaving no material  available for fallback. 
\end{description}  
These three boundaries can be derived analytically from the geometry of the debris ellipse relative to the plunge and escape boundaries.

To completely disrupt the star and produce a luminous flare, a TDE must have stellar CM orbital parameters lying between the blue and purple curves in Figure~\ref{fig:5a}. 
Among these TDEs, those lying outside the red shaded region (outside the red boundaries) will show some degree of mass deficit, because due to plunge or
escape,  some debris will be lost.  
Figure~\ref{fig:5b} shows an example in
which we calculate the fraction of stellar debris that can actually fallback.
We have assumed a solar-type star ($n$\,$=$\,$3$) disrupted by a near-extremal SMBH
($a_\bullet$\,$=$\,$0.998$) with mass $M_\bullet$\,$=$\,$6.33\times10^8\,M_\odot$. This mass is
near $M_{\bullet,\max}$, where the TDE boundary $L_{\rm TDE}$ is particularly
close to the plunge boundary $L_{\rm sep}$ (notice the logarithmic scale of the
ordinate).

The key feature shown in Figure~\ref{fig:5b} is that the fallback fraction is
strongly suppressed for nearly parabolic encounters near the plunge boundary.
For example, around the MBO, as much as $(90$--$95)\%$ of the stellar debris is
actually lost. 
In contrast, bound CM orbits behave differently: at energies
sufficiently below the parabolic limit, especially when the CM energy offset
exceeds one energy half-width $\Delta E_{\rm MBO,max}$, the debris distribution
is shifted away from the escape regions, allowing a larger fraction of material
to return to the SMBH.  This indicates that bound stellar orbits provide a
larger parameter space for producing overluminous near-horizon TDEs.

\subsection{Fallback Rate} \label{sec:3.4}

Having established the energy and angular-momentum distribution of the stellar
debris, we now calculate the fallback rates, which are conventionally used to
approximate TDE light curves.
Following the same approach as in deriving eq.~\eqref{eq: mass distribution formula }, 
we integrate over the debris distribution in the $E-L$ plane and derive the fallback rate as 
\begin{equation}\label{eq: dMdt formula}
\begin{split}
    \frac{\mathrm{d} M}{\mathrm{d} t}
    (t,\,E_c,\,L_c,\,r_t) 
    &=
    \int \!\!  \mathrm{d}\Delta E\ \mathrm{d}\Delta L \  \frac{\mathrm{d}^{2} M}{\mathrm{d}\Delta E\, \mathrm{d}\Delta L} \
    \\
    &~\times
   ~\delta \! \left(t-t_{\rm total}(E_c+\Delta E,L_c+\Delta L) \right),
\end{split}
\end{equation}
where $t_{\rm total}(E,L)$ is the elapsed time in BLC for a debris element to
travel from the tidal radius $r_t$ to the pericenter $r_p$ plus one
complete radial period $T_r$~\cite{Glampedakis2002}.  
The last equation assumes
simultaneous release of all debris elements from the star at $r$\,$=$\,$r_t$. 
In reality, the finite stellar size introduces small differences in the release
positions and coordinate times, but these corrections are negligible for
$R_\star$$\ll$\,$r_g$ and are therefore omitted.

To explore a wide range of stellar orbits, we parameterize the CM energy and angular momentum as  
\begin{subequations}\label{eq: N_E N_L formula}
    \begin{align}
        E_c & = c^2 + \hat{\eta}_E \cdot   \Delta E_{\rm MBO,max},  \label{eq:N_E formula}    \\
        L_c & = L_{\rm sep}(E_c) + \hat{\eta}_L  \cdot \Delta L_{\rm loss}(E_c),   \label{eq:N_L formula}  
    \end{align}
\end{subequations} 
where $\Delta E_{\rm MBO,max}$ is the half-width of the energy spread of the debris from an MBO TDE
(see eq.~\ref{eq: Delta E_max}), $\Delta L_{\rm loss}(E_c)=L_{\rm loss}(E_c)-L_{\rm sep}(E_c)$ is the angular-momentum gap between the boundary for complete tidal disruption and the plunge boundary,
and $(\hat{\eta}_E, \hat{\eta}_L)$ are two dimensionless parameters that specify the CM orbital energy and angular momentum relative to the chosen reference boundaries.
Notice that both $\hat{\eta}_E$ and $\hat{\eta}_L$ could be negative in this parameterization.

The resulting fallback rates for different pairs of $(\hat{\eta}_E, \hat{\eta}_L)$ are shown in Figure~\ref{fig:6}. They show two notable features.  
First, the fallback rate is highly sensitive to the CM energy.  
As $E_c$ decreases below the parabolic limit ($\hat{\eta}_E=0$), the stellar debris becomes more bound, leading to shorter
fallback times, higher peak rates, and faster declines.  For example, when $\hat{\eta}_L=1$,
the $\hat{\eta}_E=-2.4$ and $-3.6$ cases have peak rates about three orders of
magnitude higher and decay times about three orders of magnitude shorter than
the $\hat{\eta}_E=0$ case.  Second, the CM angular momentum mainly affects the
peak fallback rate.  Remarkably, while reducing $\hat{\eta}_L$ from $1$ to $0$
lowers the peak rate by merely $\sim(10$--$50)\%$, a negative $\hat{\eta}_L$
causes a suppression of the peak fallback rate by several orders of magnitude. 
The origin of this suppression lies in the stellar density profile: the debris mass is strongly concentrated toward the stellar core, and hence a substantial fraction of the debris mass is concentrated in a narrower region of the $(E,L)$ plane.

\section{Observational Implications} \label{sec:4} 

\subsection{ASASSN-15lh} \label{sec:4.1}

ASASSN-15lh is identified as a TDE based on its host-galaxy properties (a massive passive galaxy with an old stellar population) 
and its extreme peak luminosity of $L_{\rm peak}$\,$\sim$\,$(2\text{--}3)\times10^{45}\,\rm erg\,s^{-1}$~\cite{Dong2016,Leloudas2016,Brown2016,Godoy2017,Margutti2017,Kruhler2018,Maund2020}.   
The inferred SMBH mass of
$M_\bullet$\,$=$\,$5_{-3}^{+8}\times10^{8}M_{\odot}$~\cite{Kruhler2018} places it close to the maximum SMBH mass $M_{\bullet,\max}$ capable of producing an observable TDE.

However, conventional TDE models struggle to explain some of its extreme observational properties. The main difficulty lies in the light-curve evolution. In the standard parabolic-orbit picture, the fallback rate follows $\dot{M}_{\rm fb}$\,$\propto$\,$ t^{-5/3}$~\cite{Rees1988,Lodato2009}, with the characteristic fallback timescale $t_{\rm min}$ set by the return time of the most-bound debris. Since $t_{\rm min}$\,$\propto$\,$\Delta E^{-3/2}$ and $\Delta E$\,$\simeq$\,$ GM_\bullet R_\star/r_t^2$~\cite{Rees1988,Evans1989}, the inferred SMBH mass implies $t_{\rm min}$\,$\sim$\,$10^3$ days, far longer than the rapid decline observed over $\sim70$ days.

\begin{figure*}    
    \centering
    \setlength{\abovecaptionskip}{2pt}
    \vspace*{-5pt}
    \makebox[\textwidth][r]{  
        \includegraphics[width=1.025\textwidth]{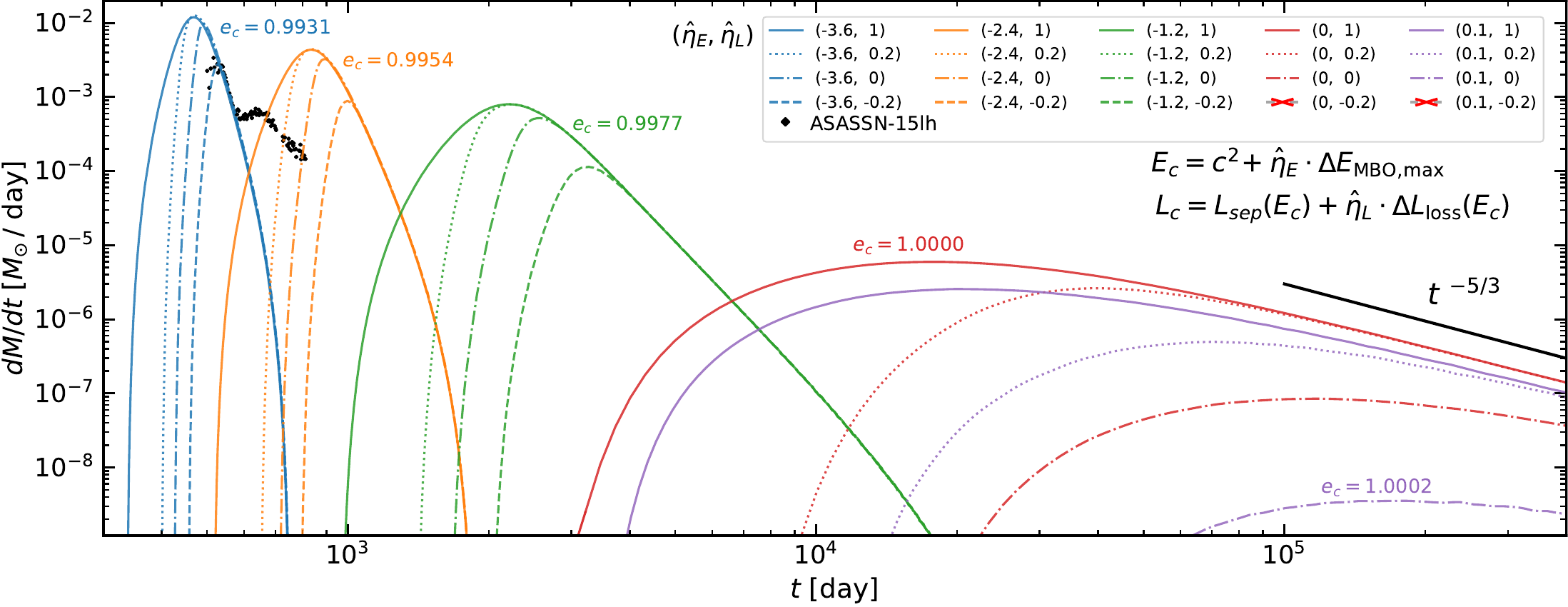}%
    } 
    \caption{
    Mass fallback rates $\mathrm{d}M/\mathrm{d}t$ for near-horizon TDEs with different orbital energies  (parameterized by $\hat{\eta}_E$) and angular momenta ($\hat{\eta}_L$). 
    The stellar mass and SMBH parameters are the same as in Figure~\ref{fig:5b}.
    The colors indicate different energies, while line styles correspond to different angular momenta.
    The solid curves mark the plunge boundary  ($\hat{\eta}_L$\,$=$\,$0$), with the CM orbital eccentricities $e_c$ indicated.
    The cases $(\hat{\eta}_E,\hat{\eta}_L)=(0,-0.2)$ and $(0.1,-0.2)$ are not shown because they produce no fallback material.
    The black diamonds show the fallback rates inferred from the bolometric luminosity of ASASSN-15lh, assuming a radiative efficiency of $\eta_{\rm eff}$\,$\simeq$\,$0.0375$.
    The time  $t$ measures the elapsed time after the star crosses the tidal radius $r_t$, so that  $t=0$ coincides with the moment of complete tidal disruption.
   } 
    \label{fig:6}    
    \vspace*{-10pt} 
\end{figure*}

One possible way to alleviate this discrepancy is to invoke relativistic effects,
which can enhance the debris energy spread and thereby shorten the fallback timescale.
However, relativistic enhancements to $\Delta E$ by a factor of
$\sim$\,$2$--$3$ via spin-orbit coupling and stellar compression
\cite{Stone2012,Kesden2012b,Leloudas2016} shorten this timescale by only a factor of a few.  
Not only is this mechanism insufficient to explain the observed fast decline, but a shortened fallback timescale would also fail to reproduce the observed peak luminosity.   
Besides, alternative models including superluminous supernovae~\cite{Kozyreva2016,Bersten2016,Chatzopoulos2016,Dai2016} and SMBH binary scenarios~\cite{Coughlin2017} also face significant challenges.

Our near-horizon TDE model may resolve this tension because of the following two features.
(i) Near-horizon TDE allows a bound stellar orbit besides parabolic ones,
which can produce shorter-period debris and lead to faster fallback.  (ii)
More importantly, a fraction of the bound debris can lie below the plunge boundary.
Such debris can be directly swallowed by the SMBH, producing a plunge-induced mass deficit
that suppresses the peak rate while preserving the rapid decline. 

To be more quantitative, we adopt $M_\bullet$\,$=$\,$6.33\times10^8M_\odot$ and
$a_\bullet$\,$=$\,$0.998$, and compare the observed light curve with our theoretical predictions.  
To convert the luminosity of ASASSN-15lh into a mass fallback rate, we use the
relation $L_{\rm bol}$\,$\simeq$\,$\eta_{\rm circ}\eta_{\rm rep}\dot M_{\rm fb}c^2$,
where $\eta_{\rm circ}$ is the fraction of rest energy released during the
collision and circularization of the fallback
material~\cite{Piran2015,Dai2015,Lu2020,Ryu2023}, and $\eta_{\rm
rep}$\,$\simeq$\,$0.25$ is the typical fraction of the above energy that is reprocessed into radiation for the inferred SMBH mass~\cite{Bonnerot2020}.  
In our model, the circularization efficiency for each debris element is given by $\eta_{\rm circ}(E,L)$\,$=$\,$(E-E_{\rm circ}(L))/c^2$, where $E_{\rm circ}(L)$ denotes the specific energy of a circular orbit with the same angular momentum as the debris element. 
Noticing that the energy and
angular-momentum spreads of the stellar debris  in near-horizon TDEs are much smaller than the corresponding CM values, $E_c$ and $L_c$, the circularization efficiency varies only weakly across the debris distribution.   
We therefore approximate the circularization efficiency by its value at the MBO, $\eta_{\rm
circ}$\,$\simeq$\,$1-E_{\rm circ}(L_{\rm MBO})/c^2$, which gives $\eta_{\rm circ}$\,$\simeq$\,$0.15$ when $a_\bullet$\,$=$\,$0.998$ (see Figure \ref{fig:3}). 
We will use this value in our later analysis. 
Finally, the effective radiative efficiency is $\eta_{\rm
eff}$\,$:=$\,$\eta_{\rm circ}\eta_{\rm rep}$\,$\simeq$\,$0.0375$.

The resulting fallback rate inferred from the light curve of ASASSN-15lh is
shown in Figure~\ref{fig:6} as the black diamonds.  
Remarkably, the peak agrees  well with what is expected from a near-horizon TDE with the parameters
$(\hat{\eta}_E,\hat{\eta}_L)$\,$=$\,$(-3.6,-0.2)$. 
In particular, the short decay timescale and the limited fallback rate, which are difficult to reconcile in other scenarios, are simultaneously satisfied in our model.  
The negative value of $\hat{\eta}_L$ indicates that the stellar CM orbit lies slightly inside the separatrix. 
Such configurations are not expected to be rare
within the near-horizon TDE parameter space, because the disruption boundary itself approaches the plunge boundary in this regime.

Notice that the late-time plateau phase observed in ASASSN-15lh is not
expected to be directly reproduced by the fallback rate alone. It likely
involves subsequent evolution of the disrupted debris and its radiation, 
including the evolution of the accretion flow and changes in the reprocessing of the central emission~\cite{Leloudas2016,Margutti2017,Mummery2020,Mummery2024b,Guo2025}.
Modeling these late-time processes would be a valuable extension to complete the light-curve analysis, but lies beyond the scope of the present work.

In addition to the plunge-induced mass deficit discussed above, partial
disruption may also reduce the fallback mass, as a fraction of the stellar
material escapes the gravitational binding of the
star~\cite{Guillochon2013,Coughlin2019,Miles2020,Guo2025}.  This effect could
also suppress the fallback rate in extreme relativistic encounters, potentially
complicating the interpretation of observed TDEs.  However, no clear signatures
of a surviving stellar core, such as recurrent
flares~\cite{Coughlin2019,Wevers2022,Liu2024}, have been identified in
ASASSN-15lh or other high-mass SMBH TDE candidates.  
In any case, explaining the short fallback timescale of ASASSN-15lh with partial disruption still requires the stellar encounter to occur in the strong-gravity regime. 
Therefore, ASASSN-15lh remains a promising candidate for probing near-horizon dynamics.

\subsection{Diagnostics for Detecting Near-horizon TDEs} \label{sec:4.2}

\begin{figure}[H] 
\centering
\setlength{\abovecaptionskip}{4pt}
\vspace*{-3pt}
\makebox[\columnwidth][r]{  
    \includegraphics[width= 1.05\columnwidth]{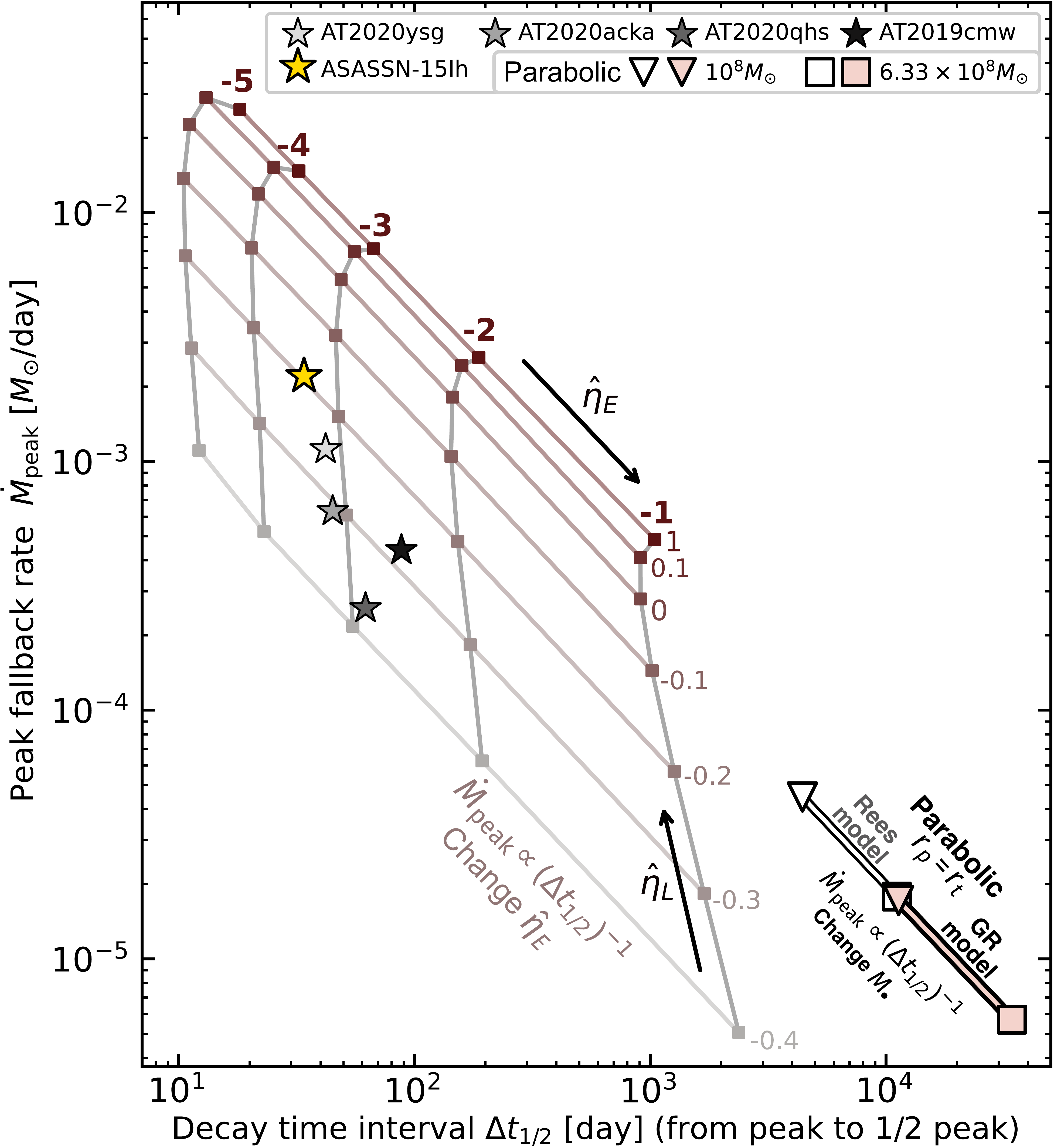}%
}   
\caption{ 
    Relationship between the peak fallback rate and the decay time interval for near-horizon TDEs. 
    Here, $\Delta t_{1/2}$ denotes the time required for the fallback rate curve to decrease from its peak value to half of the peak value. 
    The grid shows our model predictions for different  $(\hat{\eta}_E,\hat{\eta}_L)$, where we have assumed the same stellar and SMBH parameters as in Figures~\ref{fig:5b} and~\ref{fig:6}.
    The stars denote the observed high-mass TDE candidates. 
    In the lower-right part of the plot, the colored line segment connects the relativistic predictions for parabolic orbits with $r_p$\,$=$\,$r_t$ for SMBHs of masses $6.33\times10^8M_\odot$ (filled square) and $10^8M_\odot$ (filled triangle), where $r_t$ is the relativistic tidal radius. The hollow square and triangle show the corresponding standard Rees model predictions~\cite{Rees1988,Lodato2009} for parabolic orbits with $r_p=r_t$, where $r_t$ is given by the Newtonian tidal radius.  
   } 
    \label{fig:7}  
    \vspace*{-5pt}
\end{figure}

To investigate whether the near-horizon scenario applies to the TDEs other than
ASASSN-15lh, we examine in Figure~\ref{fig:7} the theoretical relationship
between the peak fallback rate and the decay timescale, and compare it with the
observed near-horizon TDE candidates~\cite{Leloudas2016,Yao2025}.  The grid  shows our model
predictions for different stellar orbital energies and angular momenta. The stellar and SMBH properties, as well as the other parameters, are the same as those adopted in sect.~\ref{sec:4.1}. 
Note that decreasing the SMBH mass is expected to shift the model grid towards the upper-left, as indicated by the position of the filled triangle relative to the filled square.

Our model prediction exhibits two interesting features.  First, for a fixed
value of $\hat{\eta}_L$, varying $\hat{\eta}_E$ produces an approximately
inverse relation between the peak fallback rate and the decay time interval, \textit{i.e.}, $\dot{M}_{\rm peak}$\,$\propto$\,$\Delta t_{1/2}^{-1}$.  
This behavior indicates that
the mass deficit mentioned in the previous sections is not sensitive to the
$\hat{\eta}_E$ parameter. This is understandable because $\hat{\eta}_E$ alone
dose not directly determine a plunge orbit. 
Second, as $\hat{\eta}_L$ decreases
to negative values, the peak fallback rate decreases significantly, and the
dependence on $\hat{\eta}_L$ is non-linear.  
The non-linearity comes from the
inhomogeneity of the stellar internal density.

To place other TDEs in Figure~\ref{fig:7}, we follow the procedure described in
sect.~\ref{sec:4.1}. 
We first convert the observed light curves into approximate
fallback rates, and then extract the corresponding peak fallback rates and decay time intervals. 
It turns out that  our model predictions agree not only with ASASSN-15lh~\cite{Leloudas2016}, but also with several other high-mass TDE candidates, whose inferred SMBH masses are
mainly concentrated around $(1$--$2)\times10^8M_\odot$~\cite{Yao2025}.

In contrast, the standard Rees
model  predicts decay time intervals of $4\times10^3$--$10^4$ days for SMBH masses between $10^8M_\odot$ and $6.33\times10^8M_\odot$, which are substantially longer than the observed values. 
Here, the debris energy distribution is obtained from the tidal disruption of the $n=3$ stellar profile, and the fallback rate is calculated using Newtonian orbital dynamics~\cite{Lodato2009}. 
Moreover, although both models follow an inverse relation $\dot{M}_{\rm peak}\propto\Delta t_{1/2}^{-1}$, the decay time interval in our near-horizon model are typically a factor of $2$--$3$ longer than those from Newtonian calculation due to extended radial orbital periods caused by strong apsidal precession, highlighting  the necessity of the near-horizon framework developed in this work.

\section{Discussion and conclusion} \label{sec:5}

In this work, we have investigated a class of TDEs in which the stellar disruption occurs
near the plunge boundary of  highly spinning SMBHs, with the tidal-disruption 
radius approaching the event-horizon scale. 
We refer to such events as near-horizon TDEs.  
Unlike conventional deep-relativistic TDEs, which typically satisfy $r_t$\,$ \gg $\,$r_p $\,$\gtrsim$\,$ r_g$, near-horizon TDEs are characterized by $r_t $\,$\sim $\,$ r_p $\,$\sim$\,$r_g$.

Based on the orbital structure in Kerr spacetime, we mapped the debris
distribution into the conserved energy--angular momentum space. We found a plunge-induced
mass deficit in the fallback rate, which can potentially be used to identify
near-horizon TDEs.  This
effect originates from the finite energy and angular momentum spread of the
debris relative to the CM orbit.  We quantify this mass deficit in
sect.~\ref{sec:3} and show that it occurs only within a specific range of CM
orbital parameters, where the debris distribution overlaps with the plunge
boundary (Figure~\ref{fig:5}). 
The effect is most significant for bound CM orbits. For nearly parabolic encounters, in contrast, a large fraction of the debris can enter the unbound region and escape the system, so that only a small fraction of the disrupted material remains available for fallback.

We further investigate the fallback rate, with particular attention to the
plunge-induced mass deficit.  For a star initially on a bound near-horizon
orbit, the intrinsically short fallback timescale is accompanied by a
suppressed peak fallback rate, as a fraction of the debris is removed by direct
plunge.  This combination of a rapid fallback and a reduced peak fallback rate
is consistent with the behavior required to explain the observed light curve of
ASASSN-15lh.  As shown in Figures~\ref{fig:6} and \ref{fig:7}, our framework
provides a viable explanation for the fallback behavior inferred for current
near-horizon TDE candidates, for which the conventional TDE model faces
difficulties.

Our results indicate that near-horizon TDE candidates are preferentially
produced by bound stars. Such orbits can arise from tidal capture, where a star
loses energy through tidal excitation during a close, non-disruptive
encounter~\cite{Fabian1975,Press1977,Vick2017,Cufari2023,Rizzuto2023,Lau2025,Yang2026},
or from the Hills mechanism, where the disruption of a stellar binary leaves 
one component bound to the 
SMBH~\cite{Hills1988,Yu2003,Bromley2006,Wang2022,Cufari2022a}. The prevalence of
these channels may reflect the long two-body relaxation timescale around
massive SMBHs, which suppresses the supply of stars on parabolic loss-cone
orbits.

It should be emphasized that our model relies on several idealized
assumptions, including equatorial orbits, linear FNC expansion for the debris
distribution, and synchronous debris release followed by geodesic evolution.
In addition, unlike deep-relativistic TDEs ($r_t$\,$\gg$\,$ r_p$), where the star is initially disrupted at $r$\,$\sim$\,$r_t$ well outside the relativistic whirl region, near-horizon TDEs undergo disruption within the strong-field region, with $r_t$\,$\sim$\,$r_p$\,$\sim$\,$r_g$. The star can therefore experience prolonged interaction with the SMBH during relativistic whirl motion, which may repeatedly exert tidal forcing, excite internal stellar oscillations, and modify the detailed disruption process~\cite{Zhou2025,Wang2026}.
Extensions to inclined orbits, higher-order tidal couplings, and  general-relativistic hydrodynamic   
simulations are therefore required for a more complete description.
Nevertheless, these effects are unlikely to alter our main conclusion that the
overlap between the debris distribution and the plunge boundary is an intrinsic
feature of near-horizon disruption. 
Future work will enable us to identify high-mass, high-spin SMBHs through their TDE signatures, thereby pinpointing more systems  suited for testing strong-field gravity.

\Acknowledgements{
We thank  Yuhan Yao for useful discussions.
This work is supported by the National
Natural Science Foundation of China (Grant No. 12473037). 
}

\InterestConflict{The authors declare that they have no conflict of interest. }


\bibliographystyle{scpma} 
\bibliography{2026SCGE}
 
\end{multicols}
\end{document}